\documentclass[10pt]{article}

\usepackage{graphicx}
\usepackage{amssymb,amsfonts,amsmath}

\usepackage{cite}

\usepackage{color} 

\usepackage[labelfont=bf,labelsep=period,justification=raggedright]{caption}

\date{}

\usepackage{marginnote}

\newcommand{\gate}{\vdash\!\dashv}

\begin{document}

\begin{flushleft}
{\Large
\textbf{Transcriptional regulation: Effects of promoter proximal pausing on speed, synchrony and reliability }
}
\\
Alistair N. Boettiger$^{1,\ast,\dagger}$,
Peter L. Ralph$^{2,3,\dagger}$, 
Steven N. Evans$^{3,4,\dagger}$
\\
\bf{1} {Biophysics Graduate Group and Department of Molecular and Cellular Biology, University of California, Berkeley, USA}
\\
\bf{2} {Department of Evolution and Ecology, University of California, Davis, USA} 
\\
\bf{3} {Department of Statistics, University of California, Berkeley, USA}
\\
\bf{4} {Department of Mathematics, University of California, Berkeley, USA}
\\
$\ast$ E-mail: Corresponding alistair@berkeley.edu
\\
$\dagger$ These authors contributed equally
\end{flushleft}

\begin{abstract} 
  Recent whole genome polymerase binding assays in the {\it Drosophila} embryo have shown that a large
proportion of unexpressed genes have pre-assembled RNA pol II transcription initiation complex stably
bound to their promoters \cite{zeitlinger2007}.  
These constitute a subset of promoter proximally paused genes 
which are regulated at transcription elongation rather than at initiation,
and it has been proposed
that this difference allows these genes to both express faster and
achieve more synchronous expression across populations of cells, thus
overcoming the molecular ``noise'' arising from low copy number factors.
Promoter-proximal pausing is observed mainly in metazoans, in accord with its posited role in synchrony. 
Regulating gene expression by controlling release from a promoter paused state instead of by regulating access of the polymerase to the promoter DNA can be described as a rearrangement of the regulatory topology so that it controls transcriptional elongation rather than transcriptional initiation.  
It has been established experimentally that genes which are regulated
at elongation tend to express faster and more synchronously; however, it has not been
shown directly whether or not it is the change in the regulated step {\em per se} that causes
this increase in speed and synchrony.
We investigate this question by proposing and analyzing a continuous-time Markov chain
model of polymerase complex assembly regulated at one of two steps: initial polymerase association with DNA, or release from a paused, transcribing state.  
Our analysis demonstrates that, over a wide range of physical parameters,
increased speed and synchrony are functional consequences of elongation control.  
Further, we make new predictions about the effect of elongation regulation
on the consistent control of total transcript number between cells, and identify which
elements in the transcription induction pathway are most sensitive to
molecular noise and thus may be most evolutionarily constrained.  
Our methods produce symbolic expressions for quantities of interest with
reasonable computational effort and can be used to explore the interplay between interaction
topology and molecular noise in a broader class of biochemical networks.
 We provide general-purpose code
implementing these methods.

\end{abstract}



\section*{Author Summary}
Gene activation is an inherently random process because numerous diffusing
proteins and DNA must first interact by random association before transcription can begin.
For many genes the necessary protein--DNA associations only begin after activation, but it has recently 
been noted that a large class of genes in multicellular organisms can assemble 
the initiation complex of proteins on the core promoter prior to activation.  
For these genes, activation merely releases polymerase from the preassembled complex to transcribe
the gene.  It has been proposed on the basis of experiments that such a
mechanism, while possibly costly, increases both the speed and the synchrony of
the process of gene transcription.  
We study a realistic model of gene transcription, and show that this conclusion 
holds for all but a tiny fraction of the space of physical rate parameters
that govern the process. 
The improved control of cell-to-cell variations afforded by regulation through 
a paused polymerase may help multicellular organisms achieve the high degree of 
coordination required for development.  Our approach has also generated tools with
which one can study the effects of analogous changes in other molecular networks
and determine the relative importance of various molecular binding rates to 
particular system properties.

\section*{Introduction}

Investigations in yeast \cite{keaveney1998,ptashne1997} led to the hypothesis that in most organisms 
the recruitment of polymerase to the promoter is the primary regulated step in the activation of gene expression \cite{juvengershon2008,Margaritis2008,Gilmour2009,Chiba2010}.
However, recent studies of multicellular organisms have revealed a
diverse array of other regulatory strategies, including several types of post-initiation regulation \cite{zeitlinger2007,muse2007,hargreaves2009}.  
Zeitlinger et al.\ \cite{zeitlinger2007} generated tissue-specific whole-genome polymerase binding data in {\it Drosophila}
and showed that regulation of polymerase release from the promoter is widespread during development.  
Their data shows that some 15\% of tissue-specific genes bind polymerase to their promoters in \emph{all} tissues, 
even though each gene only allows polymerase to proceed through the coding sequence in a specific tissue (see Supplemental Figure S1).  
Differential expression of these genes is made possible by a {\em paused state}
wherein a polymerase remains stably bound but precisely stopped a short distance from the promoter and
awaits a regulated release that is only triggered in the appropriate tissue \cite{zeitlinger2007}.   
Finally, many metazoa have been shown to have, genome-wide, disproportionate amounts of polymerase 
bound at promoter regions as compared to coding regions \cite{core2008,guenther2007,muse2007,zeitlinger2007}.  

This mechanism has been called {\em promoter proximal pausing}.  
It should not be confused with the stochastic stalling of a polymerase as it transcribes, 
a phenomenon which has also been termed ``polymerase pausing''.  
Furthermore, there are distinctions to be made between:
{\bf stalled polymerase},
a polymerase which associates in a transient, unstable manner with the promoter but does not proceed into productive transcription; 
{\bf poised polymerase},
a polymerase for which the association is stable but has not escaped from the promoter to begin transcription; and
{\bf promoter proximal paused polymerase},
a polymerase that completely escapes from the promoter but ``pauses'' in a stable, inducible state just downstream of the promoter. 
It is believed that most genes which have polymerase bound to their promoters in all tissues 
but expressed in only some tissues fall in the last category;
this promoter proximal accumulation of pol II may indicate that regulation of pausing transitions is a general feature of metazoan transcriptional control.  
We remind the reader that a gene need not use the paused state as a waiting step 
at which to integrate regulatory information in order to be termed a paused gene, 
as even constitutive house-keeping genes may be denoted as paused \cite{gilchrist2010pausing}.  
In this study we will be interested only in the elongation regulated subset of paused genes.
For further discussion of terminology and assays which distinguish these conditions, see the Supporting Information.  


It remains an open question why expression of some genes is controlled further downstream than others.  
Several groups have postulated that pausing may ready a polymerase
for rapid induction \cite{core2008,muse2007,hendrix2008}. 
(Here {\em induction} refers to the first time at which all the components 
required for expression of a particular gene become available, 
and {\em expression} is when transcription of the first nascent mRNA transcript begins.)  
To motivate this idea, the preloaded, paused polymerase is described as a ``loaded gun'' 
ready to shoot off a single transcript as soon as it is induced.  
Experiments with heat shock genes 
-- the first class of genes for which paused promoters were identified -- 
show evidence of rapid induction consistent with this idea \cite{yao2007,rasmussen1993}.   
However, pre-loading only provides an argument for
why the {\em first} transcript would be produced more quickly.  
Surprisingly then, it was also observed by Yao et al.\ \cite{yao2007} 
that {\em subsequent} polymerases are recruited rapidly to promoters of induced, elongation-regulated genes 
as well as the first, preloaded Pol II -- a phenomenon not accounted 
for by the loaded gun metaphor.  
Since most genes must be transcribed several times in order to produce functional
levels of mRNA, changes in speed of induction as a whole are likely to be of 
more physiological consequence than changes in the time at which the first, 
pre-paused transcript releases.  

When whole-genome studies extended the observation of pausing to cover many key developmental regulatory genes \cite{zeitlinger2007}, 
further questions arose.  
While the selective advantage of rapid induction is reasonably apparent for stress response genes, 
it is harder to explain why rapid induction would be selected for 
in so many developmental transcription factors and signaling pathway components.
An additional hypothesis, suggested by Boettiger and Levine \cite{boettiger2009synchronous}, is that regulation of transcriptional
elongation (for instance, by promoter proximal polymerase pausing) may have evolved to
ensure more coordinated expression across populations of cells.
This hypothesis was motivated by the
striking correspondence between genes shown experimentally to activate in a
synchronous fashion and genes shown to bind polymerase at the promoter
independent of activator state but not continue elongation until activator
arrival.   

Recent work by Darzacq and colleagues \cite{darzacq2007} provides insight into why a 
regulatory interaction downstream of transcriptional pre-initiation complex (PIC) assembly 
may lead to more coordinated gene expression than does regulation upstream of PIC assembly.  
Using fluorescently tagged transcription components,
they demonstrated that transcriptional initiation is a highly variable process,
with only about one in ninety Pol II--gene interactions leading all the way to
productive mRNA elongation \cite{darzacq2007}.  Nonproductive interactions
each lasted between several seconds and a minute, suggesting that
abortion of transcriptional initiation can occur at different stages in assembly of the complex.   
Regulatory interactions that occur after this
noisy assembly process would act only on transcriptionally competent polymerases,
and so this mechanism might result in more synchronous expression -- a hypothesis we test here.

The idea that gene expression itself is intrinsically variable (rather than variable
as a result of extrinsic fluctuations in upstream quantities) is well established and is a recent focus of
theoretical and experimental interest -- see \cite{raj2008review} and \cite{raj2009singlemolecule} for reviews.
Stochasticity can arise at many stages of the process, including from
the diffusion of molecules in the cell \cite{vanZon2006diffusion},
noisy gene regulation \cite{peccoud1995markovian},
chromatin and other conformal rearrangements \cite{degenhardt2009populationlevel},
random events during elongation \cite{rajala2010transcriptionalpausing,ribeiro2009delayed}, 
and random dynamics of translation and degradation of mRNA and proteins \cite{ribeiro2010stochastic}.

Populations of single-celled organisms have been shown to take advantage of
noisy gene expression to achieve clonal yet phenotypically heterogeneous
populations \cite{maamar2007}.  
In metazoan development, however, proper growth and development generally relies on coordination and
synchrony rather than stochastic switching.  For example, certain cells in the Drosophila embryo
are induced to become neurons if they are next to a mesoderm cell but not
mesoderm themselves \cite{derenzis2006}, so uneven activation of mesoderm fate
could produce early patches of mesoderm, thereby improperly inducing neuronal
development in neighboring tissue.  
Although synchronous behavior is important for metazoa, particularly in development, 
it is not a universal property of all metazoan genes.
For instance, genes  
with both synchronous and very stochastic patterns of induction 
 have been observed in the Drosophila embryo \cite{boettiger2009synchronous}.  
The unique challenges of coordinating the behavior of a large number of independent 
cells may explain why elongation regulation aimed at release from a paused state 
appears to be much more dominant among metazoa like {\it D.\ melanogaster} and humans 
than \emph{E.\ coli} or \emph {S.\ cerevisiae}.  

Here we investigate mathematically whether the significant change in the
coordination of expression observed in experiment \cite{boettiger2009synchronous} 
can be explained by a change in the regulation
network topology which only effects whether regulation occurs before 
or after PIC assembly, while keeping other details (reactions and rates) of the PIC 
assembly process the same.  
We also seek to determine which interactions in the transcriptional pathway are
most important for determining the coordination of expression, and what effect
different topologies have on the speed of induction and variability between 
sister cells in total number of mRNA synthesized.

\begin{figure}[!ht]
  \begin{center}
    \includegraphics[width=\textwidth]{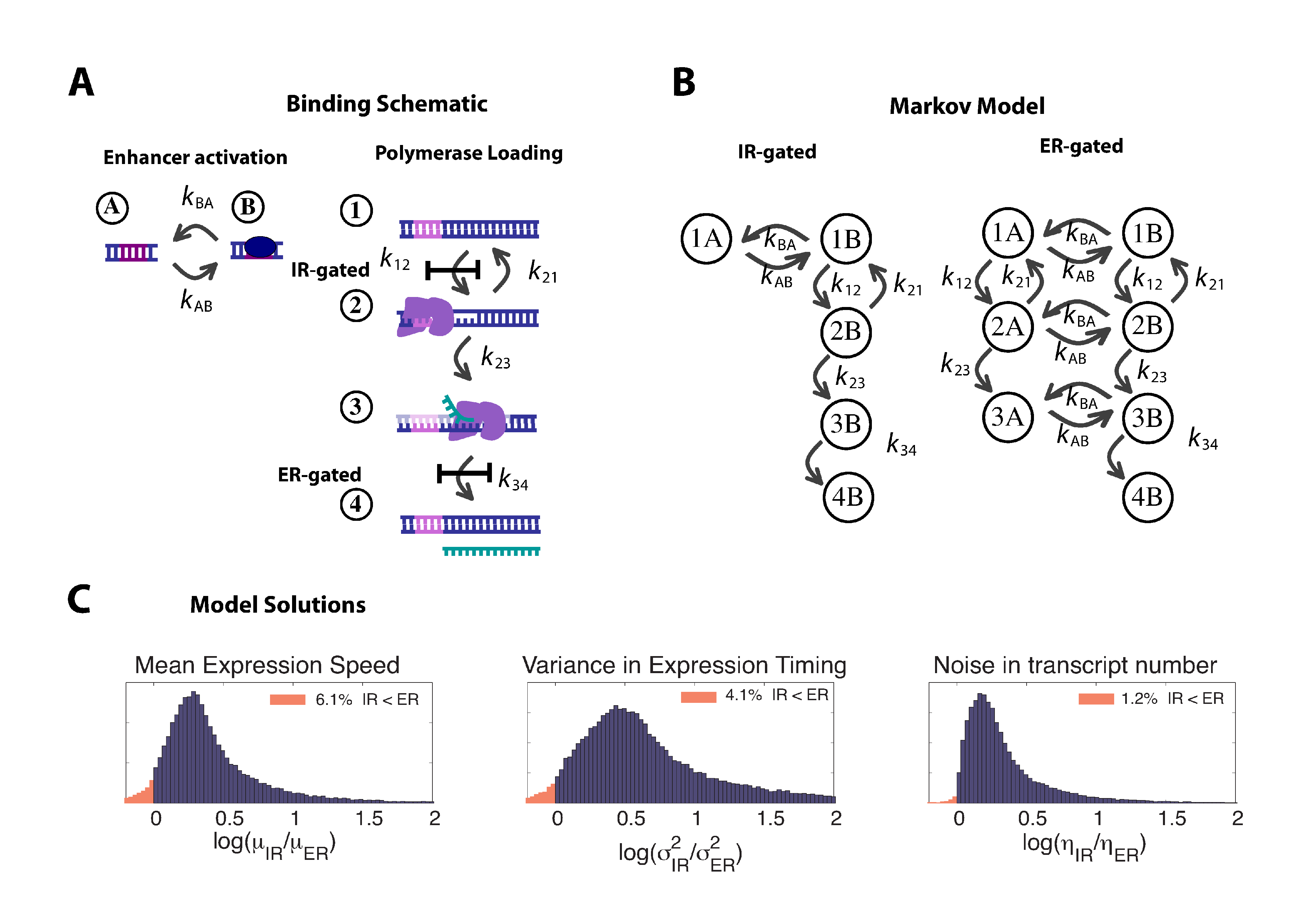}
  \end{center}
  \caption{ {\bf From regulatory mechanism to Markov Chain:} {\bf (A)} Schematics of two simplified models for initiation regulation (IR) and elongation regulation (ER). 
Transcription is represented in 4 steps: (1) naked DNA, (2) DNA-polymerase complex, (3) actively transcribing polymerase, and (4) completed mRNA.  
The enhancer is either (A) open or (B) bound.  
The enhancer must be bound (the permissive configuration) for the transcription chain to pass the gated step ($\gate$), whose identity depends on the model (IR or ER). 
{\bf (B)} The corresponding Markov chains for each regulation scheme.  
Colors of arrows denote the transition rates from (A).  
Note that one set of rate parameters determines all the numerical values for both chains, allowing for a direct test of the effects of topological change.}
{\bf (C)} Distributions of log ratios of speed ($\mu$), variance of expression time ($\sigma$), and transcript count variability ($\eta$)
across 10,000 randomly chosen parameter vectors (as described in the text), 
showing that ER is faster, less variable, and produces less variability in transcript numbers over most possible combinations of rate parameters
for this simple model.
  \label{fig:markovmodel}
\end{figure}

We do this by constructing continuous-time Markov chain models of PIC assembly
with states that correspond to joint configurations of the promoter and the enhancer.
The (random) time taken for the chain to pass from a ``start'' state to an ``end'' state 
corresponds to the elapsed time between successive transcription events.
The models we construct for the two different modes of regulation have a common set of transition rates, 
but the particular mode of regulation dictates that certain transitions are disallowed,
resulting in two chains with different sets of states accessible from the ``start'' state.
We describe this situation by saying that each model is a {\em topological rearrangement} of the other.
Because the same set of transition rates completely parametrize both chains, (see figure \ref{fig:markovmodel}) 
we can make meaningful comparisons between the two models.
Once the Markov chains are constructed, 
we use the Feynman--Kac formula \cite{fitzsimmons1999fk}, 
model-specific decomposition techniques and computer
algebra to find symbolic expressions for features of
these first passage times that correspond to the delay between induction and transcription.

Although there has recently been much work modeling different sources of stochasticity in gene expression, 
most models refrain from a detailed representation of the different protein--DNA complexes involved in favor of more abstract approximations \cite{bialek2008cooperativity,pedraza-paulsson,pedraza2005noise, Thattai2001,thattai2002attenuation,tkavcik2008rin,maamar2007}. 
Two--state ``on--off'' Markov chains have been used many times to model stochasticity in transcription 
(e.g.\ \cite{peccoud1995markovian,becskei2005lownumber}),
and provide analytic solutions.
Such models have been used to explain, for instance, the observation that mRNA copy number does not in general follow Poisson statistics,
implying that there are ``bursts'' of transcription in some sense.
This bursting behavior can occur if the gene transitions between an {\em active} state (in which transcription can occur),
and an {\em inactive} state (in which it does not), as shown by Raj et al.\ \cite{raj2006mRNAsynthesis}.
Although more complicated Markov chain models have appeared,
often presented via a stochastic chemical master equation \cite{samoilov2006deviant},
they are usually simulated rather than studied analytically 
(see \cite{resat2009kinetic} for a review of methods and software).
A notable recent exception is Coulon et al.\ \cite{coulon2010spontaneous},
who use matrix diagonalization to study the power spectrum and other properties of several models of regulation.
A complementary set of techniques takes a broader view,
using the fluctuation--dissipation theorem to
work on the scale of \emph{small} stochastic deviations 
from the differential equations that capture the average behaviors at equilibrium
\cite{bialek2008cooperativity,pedraza-paulsson,pedraza2005noise,thattai2002attenuation,tkavcik2008rin}.

We model the intrinsic noise of regulation and polymerase recruitment
using biologically-derived Markov chain models.
We focus on this particular piece of the larger process of expression 
in greater detail than has been done previously in order to
provide a detailed mathematical investigation of the role of promoter proximal pausing.
Unlike simulation methods, our approach provides a tractable way to compute analytic expressions for which interpretation is direct and reliable.  
Moreover, it does not depend on small-noise or equilibrium assumptions, or require the passage to a continuum limit.
Furthermore, the structure of the models we use is determined by biological realism
rather than being constrained by mathematical tractability.
Our approach is most similar to that of \cite{coulon2010spontaneous},
although our methods are less computationally intensive and produce symbolic expressions 
which allow us to investigate phenomena in greater depth.
In particular, we compare alternate modes of gene regulation and readily
evaluate analytically the sensitivity of system properties to changes in rate parameters 
over a large proportion of parameter space. 

\section*{Methods}

\subsection*{Framework for modeling regulatory interactions}
As a prelude to describing the actual Markov chain model of transcriptional regulation we analyze, 
we describe a general approach to modeling promoters, enhancers and their interactions, and
illustrate this approach with a toy model of transcription that is not too cumbersome to draw
-- see figure \ref{fig:markovmodel}.

We begin with two separate Markov chains, a {\em promoter chain} 
and an {\em enhancer chain} (figure \ref{fig:markovmodel}A).
The states of the promoter chain are the possible configurations of the 
components involved in polymerase loading onto the promoter
(e.g.\ ``naked DNA'' or ``DNA--polymerase complex'') 
and the allowable transitions correspond to the arrivals of these components,
in whichever order is permissible by the underlying biochemistry.
The states of the enhancer chain are the
the components involved in enhancer activation 
(e.g.\ the binding of regulatory transcription factors 
to the appropriate cis-control sequence for that promoter).  

Next, to model the regulatory interaction between enhancer and promoter, 
we designate a particular configuration of the enhancer as the {\em permissive configuration}, 
and specify a particular transition of the promoter chain as the {\em regulated step}.  
We  require the enhancer chain to be in the permissive configuration 
for the promoter chain to make the transition through the regulated step
and we assume that the enhancer remains in the permissive configuration 
as long as the promoter chain is downstream of that step.
(The specification that the enhancer remains in the bound/permissive state 
while the process is downstream of the regulated step is not the only possible choice, 
but it is perhaps the most realistic.) 
We choose the regulated step according to the regulation mechanism that
we are modeling.

The composite stochastic process that records the states of both the promoter and enhancer chains 
is our resulting Markov chain model of transcription. 
Varying the regulated step leads to alternative topologies for this chain.
We stress that, as we change the choice of regulated step, 
the underlying promoter and enhancer chains remain the same.
In particular, the same set of rate parameters are used in both schemes and they have the same meaning.
This permits meaningful comparison of different methods of regulation.
Two possible regulated steps, labeled ``IR gated'' and ``ER gated'',  
are shown  along with the corresponding Markov chains in figure \ref{fig:markovmodel}.
Each possible configuration of the components of the transcription complex 
and associated enhancer elements is represented by a state of the 
composite chain,
and the composite chain jumps from one state to another when a single molecular 
binding or unbinding event converts one configuration of complexes into another.   
For simplicity, we assume that each arrival in the end state allows one transcript to be made.
After transcription occurs, the transcription complex may dissociate entirely, 
returning the chain to its initial state,
or it may leave behind a partial {\em scaffold}, returning the composite chain to an intermediate state
(and possibly leading to successive rounds of reinitiation and thus a ``burst'' of transcription products --  i.e.\ multiple mRNA molecules being transcribed per promoter opening event). 

Formally, the general composite Markov chain model is constructed as follows.
Consider two promoter configurations, say, $x_i$ and $x_j$, such that a direct transition from
the first to the second is possible.
Write $r_P(x_i,x_j)$ for the rate at which this transition occurs.
For any two promoter configurations for which a direct transition is not possible, we set this rate equal to zero.
Similarly, we write $r_E(y_i,y_j)$ for the transition rate from enhancer configuration 
$y_i$ to enhancer configuration $y_j$.
Denote the permissive enhancer configuration by $y_*$.
Suppose that the regulated step of the promoter chain is the step from state $x_a$ to state $x_b$.
Let $X^*$ be the set of states downstream from $x_b$, 
i.e.\ those states that can only be reached from the unbound state by passing through $x_b$.
Then, the composite Markov chain takes values in a set of pairs of configurations $(x,y)$, 
and it jumps from $(x_i,y_i)$ to $(x_j,y_j)$ at rate $q( (x_i,y_i), (x_j,y_j) )$, defined as follows:
\begin{align*}
   q( (x_i,y_i), (x_j,y_i) ) &= r_P(x_i,x_j), \qquad &\mbox{if} \; (x_i,x_j) \neq (x_a,x_b), \\
   q( (x_i,y_i), (x_i,y_j) ) &= r_E(y_i,y_j), \qquad &\mbox{if} \; x_i \notin X^*,\\
   q( (x_a,y_i), (x_b,y_i) ) &= 0,            \qquad &\mbox{if} \; y_i \neq y_*,\\
   q( (x_a,y_*), (x_b,y_*) ) &= r_P(x_a,x_b), \qquad &  
\end{align*}
and $q((x_i,y_i),(x_j,y_j))=0$, otherwise.  
Denote by $x_e$ the expressing promoter configuration with productively elongating mRNA.
We are interested in the passage of the composite Markov chain from certain starting states --
either the state in which both promoter and enhancer are unbound or the state to which the system returns after elongation begins --
to the final, expressing state $(x_e, y_*)$.
Depending on which transition is regulated, some pairs of promoter and enhancer configurations will be unreachable from the relevant starting states;
these pairs are biochemically inaccessible and are never visited,
and so need not appear in our depictions or in our generator matrices (e.g.\ state 2A in the IR-gated model of figure \ref{fig:markovmodel}).

Because there are generally only two promoters per gene active at the same time
in a given nucleus, binding of a general transcription factor (TF) at one locus does not decrease the total
concentration of the TF in the nucleus sufficiently to affect the rate of
binding at the homologous locus.   Furthermore, since the observed timescales
of variability in induction are shorter than the expected timescale for protein
translation and folding, we neglect any feedback from mRNA synthesis which might modify the transition rates.  
This allows us, in particular, to assume that the jump rates of the Markov chain are homogeneous in time.

\subsection*{Detailed model of transcription}

We now apply this framework to examine a model of transcription that is more interesting and detailed 
than the toy model used above for illustrative purposes.

Many general transcription factors (TFs), such as the protein complexes TFIIA,
TFIIB, etc., function together in a coordinated fashion to form the pre-initiation complex (PIC) necessary for the proper activation of transcription \cite{hager2009transcription,kornberg2007,thomas2006}. Experiments with
fluorescently labeled TFs {\em in vivo} indicate that the components of this complex
assemble on the promoter DNA \cite{darzacq2007,sprouse2008} rather than
float freely in the nucleoplasm, as had been previously argued
\cite{parvin1998}.  

\begin{figure}[!ht]
  \begin{center}
    \includegraphics[width=4in]{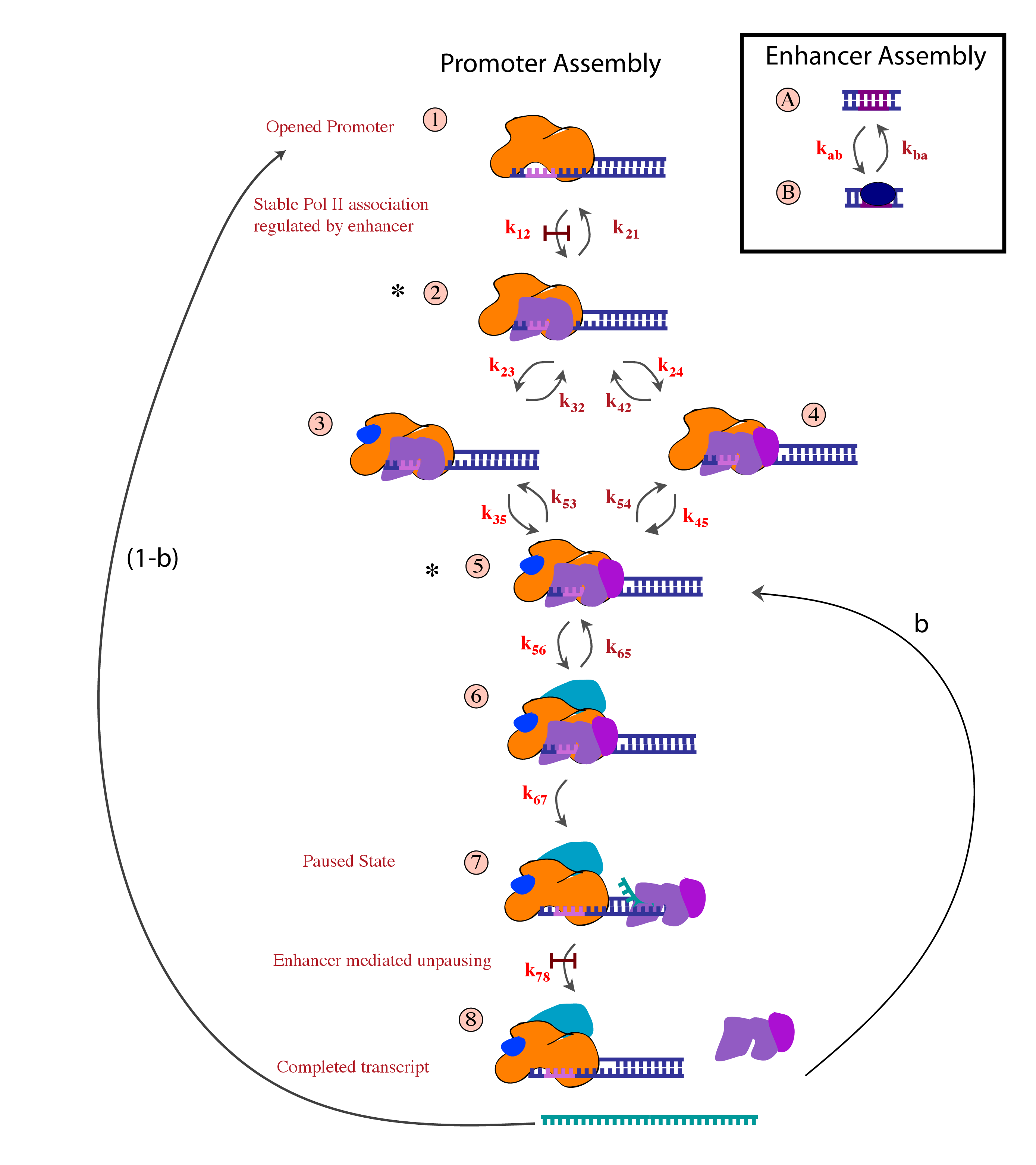}
  \end{center}
  \caption{ {\bf Model of PIC assembly.}  
Each possible complex in the process is enumerated as a state of the promoter Markov chain.  (see text for description of each complex)
The promoter chain (states 1--8) is combined with the enhancer chain (states A and B) to make the full 16 state model of transcription. 
Transitions that in some scheme require an activated enhancer (state B) are indicated by a gate, $\gate$.  
Forward rate transitions are in light font and backward transitions in dark font.  
The $1 \to 2$ transition is regulated in the IR scheme, and the $7 \to 8$ transition is regulated in the ER scheme.
}
  \label{fig:picassembly}
\end{figure}

The steps of PIC assembly are not fully understood \cite{hager2009transcription}, 
although some important details are known.  
We analyze the assembly scheme depicted in figure \ref{fig:picassembly}, which is largely consistent with available data.  
The promoter is recognized by TFIID,  the binding of which allows
TFIIA and TFIIB to join the complex \cite{thomas2006}.  
We choose this complex as the first state in our promoter model (state 1 of figure \ref{fig:picassembly}), 
since it is only just after this step that the regulation method may differ.   
TFIIB facilitates the recruitment of RNA polymerase II (Pol II) \cite{thomas2006} (state 2).  
For many non-paused genes, polymerase is only detected in cells that have an
activated enhancer (the cis regulatory sequence which controls expression) \cite{zeitlinger2007}.  
We call these genes {\em initiation regulated} and require that the enhancer 
reach its permissive state ($B$) before this association can occur.  Since Mediator is important for many
promoter--enhancer interactions \cite{hager2009transcription,fuda2009} it has
likely also joined the complex prior to polymerase arrival.  
TFIIE, (state 3), and TFIIF (state 4), bind next, possibly in either order.   
Once both are bound (state 5), TFIIH must also bind (state 6) 
before Pol II starts synthesizing RNA and clears the promoter \cite{kornberg2007, hager2009transcription}. 
TFIIH is displaced upon promoter escape \cite{kornberg2007}, 
and if Ser 2 of the Pol II tail is not phosphorylated by CDK9 (pTEFb), 
transcription pauses 40--50 base pairs downstream of the promoter \cite{rasmussen1993,fuda2009,sims2004} (state 7).  
For elongation regulated genes, it is the release from this paused state that is 
possible only in the presence of an activated enhancer (permissive state) 
-- which is generally believed to recruit the necessary CDK9 (and possibly other factors).   
Phosphorylation of Ser 2 allows the fully competent polymerase to proceed through the gene and produce a complete mRNA (state 8).  The transition rates between configurations depend on the energy of association of the bond created 
and the concentration of the reacting components.

Since we are interested in exploring the differences in which step of PIC assembly is regulated and not the different possible modes of enhancer activation, we use a simple abstracted two-state model of enhancer activation.  
A single transition switches the enhancer from the inactive state to the permissive state.  
For instance, a transition to the permissive state could represent the binding of a TF to the enhancer.
This is not likely to be completely realistic, but
if a particular step in the actual dynamics of transcription factor assembly and enhancer-promoter interaction is rate-limiting (e.g.\ the looping rate between a bound enhancer and its target promoter), then its behavior will be well approximated by our minimal model, with the transition from active to inactive corresponding to the rate for this limiting step.  

For many paused genes, it is the phosphorylation event which is believed to be
regulated \cite{zeitlinger2007,fuda2009}.  
However, accumulating data suggests
the molecular identity of the release factors may vary between paused genes.  
For example, some also require the recruitment of TFIIS in order to
escape a ``backtracked'' paused state \cite{adelman2005}.  
We consider any such regulation by release from pausing after PIC assembly to be {\em elongation regulation} (ER),
and any regulation acting upstream of PIC assembly {\em initiation regulation} (IR).

Finally, the scaffold of transcriptional machinery that facilitates polymerase binding 
does not necessarily dissociate when transcription begins. Thus, 
reinitiation may occur by binding new polymerases (at step 5)
which must still reload TFIIH which was evicted during promoter escape in order 
to proceed to step 6 and so on back to step 8.  
Repeated cycles of reinitiation may lead to a burst of mRNAs synthesized from a single promoter opening event.  
We denote by $b$ the probability that the scaffold survives to cycle in a new polymerase (see figure \ref{fig:picassembly}).
The scaffold breaks down before the next polymerase arrives
with probability $1-b$, in which case transcription activation must start again from state 1.  
We analyze both the time until the first transcript begins (for which such bursting is irrelevant)
and the effect of this partial stability of the scaffold on cell--to--cell variation in total mRNA.  

Our aim is not to present a definitive model of PIC assembly itself.
Rather, we seek to understand the impact of different modes of
regulation on a reasonable model that incorporates sufficient detail and to
develop tools that can analyze effectively models of this complexity.

\subsection*{Statistical Methods}

We are interested in the speed and variability of the transcription process,
as measured, respectively, by the mean, $\mu_{\tau}$, and variance, $\sigma^2_{\tau}$, of the delay $\tau$
between induction of the gene and expression of the first functional mRNA transcript.  
(Recall that by {\em induction} we mean the first time at which all the components 
required for expression of a particular gene become available, 
and by {\em expression} we mean the time when transcription of the first nascent mRNA transcript begins.)
We use the mean delay to explore the hypothesis that the mechanism
of elongation regulation is faster than that of initiation regulation, even when there
is no polymerase initially bound (as reported in \cite{yao2007}).
The variance of the delay is related to the degree of
synchrony of expression of the first transcripts in a population of identically
induced cells (studied in \cite{boettiger2009synchronous}) -- allowing us to test if synchrony is a functional consequence of elongation regulation.
We are also interested in the variation between activated cells of 
the total amount of mRNA produced in each.  If we denote by $N(t)$
the random number of transcripts produced up until time $t$, then it follows from
elementary renewal theory (see e.g.\ Section XI.5 in \cite{feller-vol-2}) that $N(t)$ has mean
approximately $\mu_{N(t)} \approx t/\mu_{\tau}$ and variance approximately $\sigma^2_{N(t)} \approx \sigma^2_{\tau} t/\mu_{\tau}^3$.  
A natural measure of relative variability of $N(t)$ is the squared coefficient of variation of $N(t)$, $\sigma^2_{N(t)}/\mu^2_{N(t)}$
(i.e. \ the variance of $N(t)$ divided by the squared mean of $N(t)$), which is thus approximately $\sigma^2_{\tau}/(\mu_{\tau} t)$.  
We denote the coefficient $\sigma^2_{\tau}/\mu_{\tau}$ by $\eta$, and refer to it as {\em transcript count variability}. 
The transcript count variability provides a measure of the variation in total number of rounds of transcription initiated by identical cells that have been induced for the same amount of time.  Note that $\eta$ 
has units of time:
\begin{displaymath}
  \eta = \frac{\sigma_{\tau}^2}{\mu_{\tau}} \approx \frac{\sigma_{N(t)}^2 }{\mu_{N(t)}^2}t .
\end{displaymath}
However, the ratio of this quantity for the IR scheme to its counterpart for ER scheme does not 
depend on our choice of time scale.   For any time $t$, this ratio
is approximately the ratio of the squared coefficients of variation of $N(t)$ for the two schemes,
and thus the ratio provides a way of comparing the relative variability in transcript counts between
the two schemes across all times.
Such a comparison is of interest because many of the known pausing regulated genes are
transcription factors or cell signaling components that act in concentration
dependent manners, and hence the precision of the total number of transcripts
made directly affects the precision of functions downstream \cite{boettiger2009synchronous}.  
(Rather than the coefficient of variation, some authors consider the Fano factor of $N(t)$,
defined to be $\sigma^2_{N(t)} / \mu_{N(t)}$ \cite{Thattai2001}.
If $N(t)$ has a Poisson distribution, then its Fano factor is 1,
and hence a Fano factor that differs from 1 indicates some form of ``non-Poisson-ness''.
As such, the Fano factor capture a feature of the {\em character} of the stochasticity
inherent in the number of transcripts made up to some time,
whereas the squared coefficient of variation indicates the (relative) magnitude of the stochastic effects.)

We use our model to examine how these three important system properties 
-- speed, synchrony, and transcript count variability 
-- depend on the jump rates and how they differ between an IR and an ER regulation scheme.  
In both cases, the delay $\tau$ between induction and transcription corresponds to 
the (random) time it takes for the corresponding Markov chain to go from an initial state $s$ to a final state $f$.  
For the chains corresponding to the models shown in figures \ref{fig:markovmodel} and \ref{fig:picassembly}, 
the moments of $\tau$, the Laplace transforms of $\tau$, and hence the probability distributions themselves,
can be found analytically as we describe briefly here (for detailed discussion, see the Supporting Information, Text S1; and figure S2). 

\begin{figure}[!ht]
  \begin{center}
  \includegraphics[width=6in]{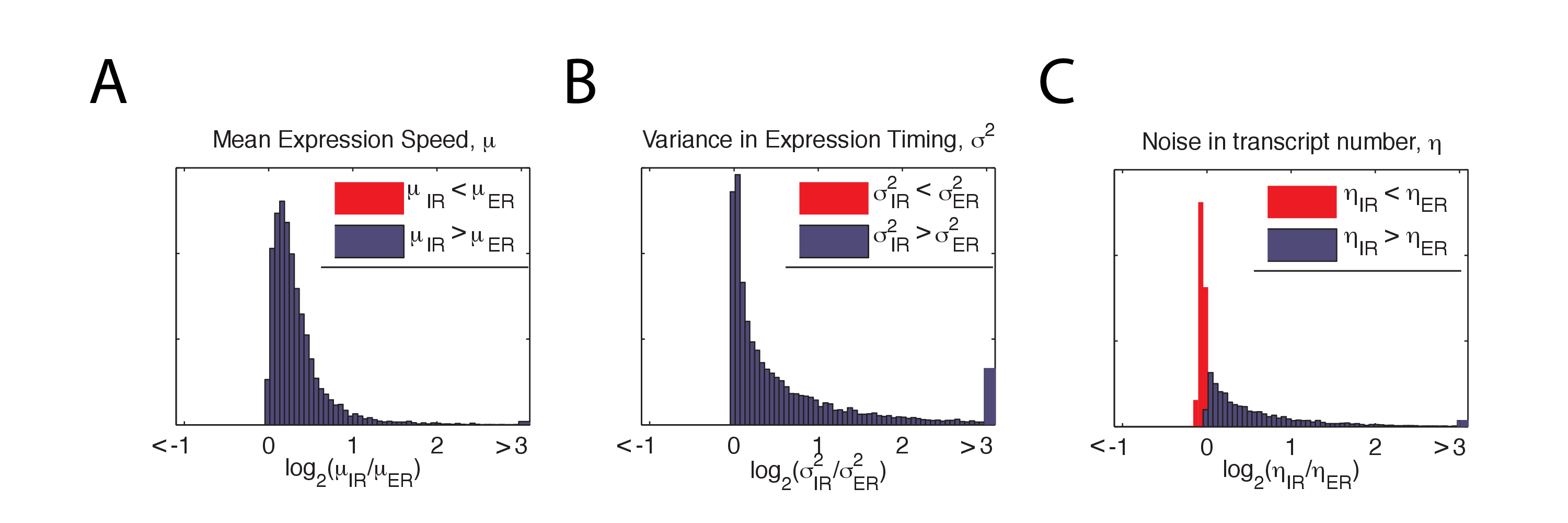}
  \end{center}
  \caption{ {\bf Model Results:} {\bf (A)} Comparison of log ratios of mean expression speed for the IR/ER schemes for 10,000 uniformly sampled rates.  For all jump rates, the log ratio is positive (red line), indicating the ER scheme is always faster.  Extreme values that would be off the edge of the graph are collected into the outermost bins.  {\bf (B)} Variance in timing of expression.     {\bf (C)} $\log_2$ ratio of noise in transcript number, measured by the squared coefficient of variation between cells of total mRNA counts $N(t)$ up to time $t$: $\sigma_{N(t)}^2/\mu_{N(t)}^2$ -- the ratio is approximately independent of $t$. }
  \label{fig:modelresults}
\end{figure}

Denote by $Q$ the {\em infinitesimal generator} matrix that has off-diagonal entries $q_{ij}$ given by the jump rate from state $i$ to state $j$, and diagonal entries $q_{ii}$ given by the negative of the sum of the jump rates out of state $i$. 
The infinitesimal generator of the chain {\em stopped when it hits state $f$} is the matrix $\widetilde Q$ 
obtained by replacing the entries in the row of $Q$ corresponding to $f$ with zeros.  
Writing $p(\cdot)$ for the probability density function of $\tau$, 
the Laplace transform of $p$ is
\begin{equation}  \label{eqn:laplacetransform}
  \phi(\lambda) = \int_0^\infty e^{-\lambda t} p(t) \, dt = (\lambda I - \widetilde Q)^{-1}_{sf} .
\end{equation}
In principle, the transform $\phi$ can be inverted to find $p$, as we do in figure \ref{fig:modelresults}D.  
Also, the $n^\mathrm{th}$ moment of $\tau$ can be found from the $n^\mathrm{th}$ derivative of $\phi$:
\begin{equation} \label{eqn:moments}
  \int_0^\infty t^n p(t) \, dt = (-1)^n \frac{d^n}{d \lambda^n} \phi(\lambda) \Big \vert_{\lambda = 0} .
\end{equation}
In particular, the mean and variance of $\tau$ can be computed from the first and second derivatives of $\phi(\lambda)$.

It is not necessary to carry out the differentiation in equation \eqref{eqn:moments} explicitly, since
\eqref{eqn:moments} becomes
\begin{equation}
  \int_0^\infty t^n p(t) \, dt = n! \sum_y \left(-Q_{-f}\right)^{-(n+1)}_{sy} \widetilde{Q}_{yf}
\end{equation}
after some matrix algebra, as derived in the Supporting Information.  
Here, $Q_{-f}$ is the submatrix of $Q$ obtained by removing the final row and column.
As shown in the Supporting Information, these expressions can be computed much more efficiently than \eqref{eqn:laplacetransform} or \eqref{eqn:moments}.

Equation \eqref{eqn:laplacetransform} is known as the Feynman--Kac formula \cite{fitzsimmons1999fk}, 
and it reduces our problem in principle to inverting the matrix $( \lambda I - \widetilde Q )$.
This is easy to do numerically for particular rate parameter values,
but in order to make detailed general predictions about the consequences 
of changing the step at which the enhancer regulates transcription
we require symbolic expressions for the system properties with the rates as free parameters.
However, for even moderately complex chains like that described in figure \ref{fig:picassembly}, 
symbolic inversion of the matrix is prohibitively difficult for commonly available software. 

To overcome this obstacle, we develop new analytic techniques 
that take advantage of the special structure of these  matrices.
First, we note that chains modeling transcription often have a block structure,
in that we can decompose the state space according to the subset of
states that must be passed through by any path of positive probability 
leading from the initial to the final state (we call such states {\em pinch points}) (see figure \ref{fig:picassembly}).  
A schematic of this decomposition is shown in figure S2.  
The models of initiation regulation we consider are amenable to this approach.  
In order for the ER model to be amenable to this approach, we assume that by the time the PIC assembly has reached the regulated step,
the enhancer chain is in (stochastic) chemical equilibrium.
Concretely, if $\pi$ is the stationary probability that the enhancer is in the permissive state,
then at each time the promoter chain jumps to state 7 (of figure \ref{fig:picassembly}) we suppose it jumps to state 7B with probability $\pi$
and to state 7A with probability $(1-\pi)$.
(To evaluate the effect of this approximation, 
we investigate how our results change after removing the parameter vectors
in which the enhancer chain is slow to equilibrate and hence
when this approximation is the worst.)
A similar decomposition for elongation regulated genes is possible using spectral theory,
but the computational savings are not as great as for the pinch point decomposition. 
We provide a detailed description of these techniques and the accompanying proofs 
(plus implementations coded in MATLAB) in the Supporting Information.

Our approach has several advantages.  Firstly, once we have derived
symbolic expressions for features of interest, it is straightforward to substitute in a
large number of possibilities for the transition rate vector in order to understand how
those features vary with respect to the values of the transition rates.  
This would be computationally impossible using simulation and at best
very expensive using a numerical version of the naive Feynman--Kac approach.
Secondly, we are able to differentiate
the symbolic expressions with respect to the transition
rate parameters to determine the sensitivity with respect to the values of the parameters.
It would be even more infeasible to use simulation or a numerical Feynman--Kac approach
to perform such a sensitivity analysis.

\section*{Results}

\subsection*{Predictions for representative parameter values}

To get an initial sense of the differences between these two schemes of regulation, 
we first compared the transcriptional behaviors for a best-guess set of parameters, 
guided by measurements of promoter binding and escape rates by Darzacq et al.\ \cite{darzacq2007} and Degenhardt et al.\ \cite{degenhardt2009} {\it in vivo} and observations in embryonic Drosophila transcription.
These data do not allow us to uniquely estimate all 14 binding reaction rates in our model of PIC assembly, 
but they do constrain key properties, including the time scale of the rate-limiting reactions 
and the ratio of forward to backward reaction rates for both early binding events and later promoter engagement events.  
We chose parameters to be consistent with these measurements,
and chose enhancer activation and deactivation rates to be
consistent with induction times estimated in Drosophila \cite{boettiger2009synchronous}
(which are also in the range recently reported in human cell lines \cite{degenhardt2009}).
%

We used the following rate parameters for the model of figure \ref{fig:picassembly}: 
\[
\begin{split}
& [ k_{12}, k_{21}, k_{23}, k_{32}, k_{24}, k_{42}, k_{35}, k_{53}, k_{45}, k_{54}, k_{56}, k_{65}, k_{67}, k_{78},k_{ab}, k_{ba}] \\
& \quad =   [.108, .725,   10,  10,  10,   10,   10,  10, 10,  .008, .005, 10, 10,  10, .01, 1] \mbox{sec}^{-1}. \\
\end{split}
\]    

\begin{figure}[!ht]
  \begin{center}
  \includegraphics[width=4in]{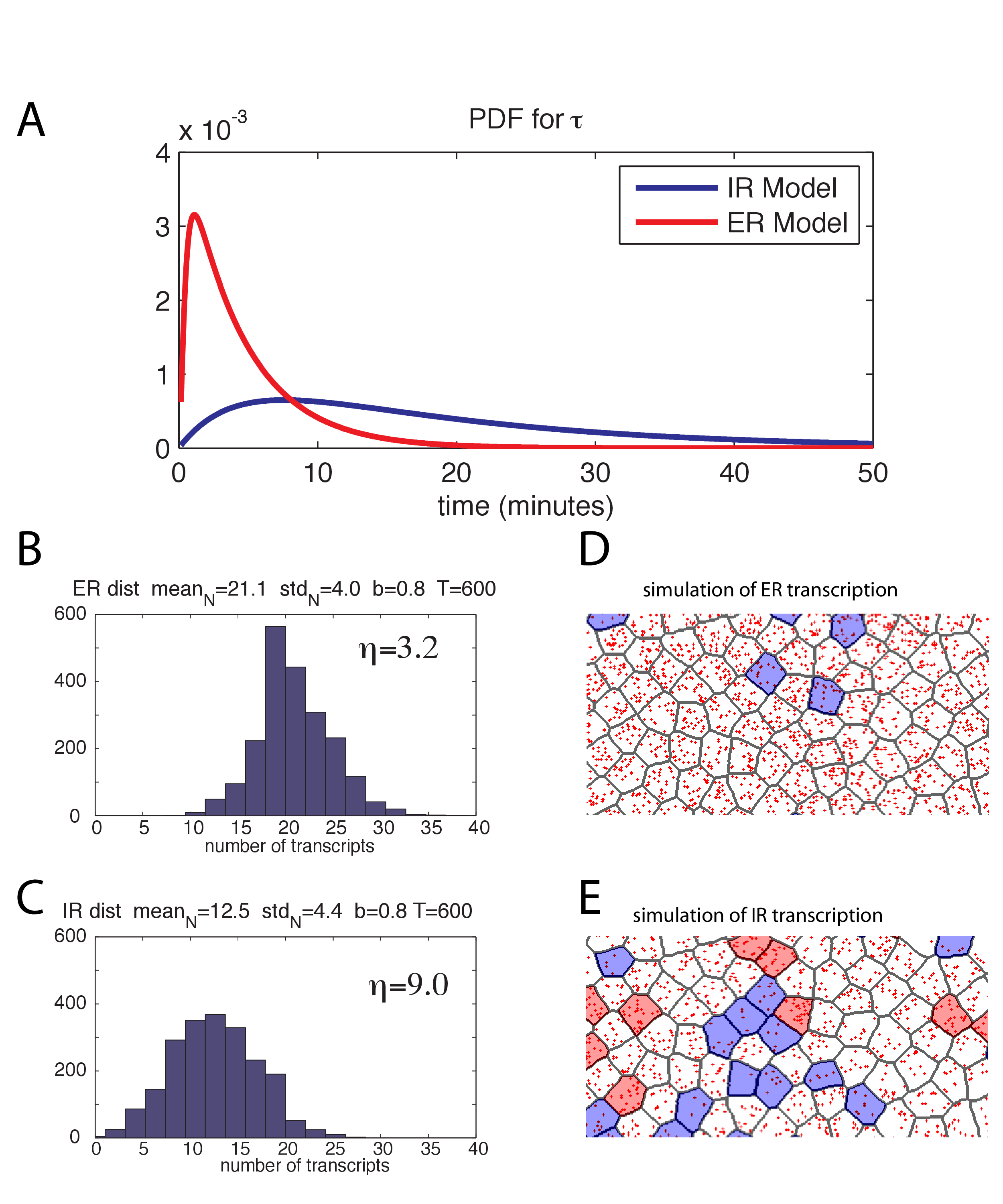}
  \end{center}
  \caption{ {\bf Model Predictions:}  {\bf (A)} Probability distributions for first passage times: Probability density functions of the time to first transcription, obtained by inversion of symbolically calculated Laplace transforms, using rate parameters computed in experimental studies of particular transcription systems.  Rates inferred from Darzacq et al.\ \cite{darzacq2007} measurements of promoter binding and promoter escape rates (see text).  
{\bf (B)} Distribution of total transcripts among a population of simulated cells during 600 minutes of transcription under the ER model with parameters as in (A) and a reinitiation probability of 0.8.  {\bf (C)} as in (B) but for the IR model.      {\bf (D)}  Individual cell simulation (see text) showing of the expected results for an mRNA counting assay on the population of cells plotted in (B).  Each mRNA transcript is represented by a red dot randomly positioned within the cell.  Cells with less than two-thirds of mean mRNA concentration are shaded blue, cells with more than three-halves of mean mRNA concentration are shaded red.   {\bf (E)}  as in (D) but for the IR scheme.}
  \label{fig:sim_results}
\end{figure}

We found the probability density of the amount of time it takes the system to go from induced to actively transcribing,
shown in figure \ref{fig:sim_results}A,
by numerical inversion of the Laplace transform 
(equation \ref{eqn:laplacetransform}).
With these rate parameters, 
the mean time between induction and the start of transcription 
for an elongation regulated scheme is around 5 minutes, 
with a standard deviation of about 4 minutes,  
whereas an initiation regulated scheme with the same rate parameters 
has a mean of 16 minutes and a standard deviation of 12 minutes, 
consistent with experimentally estimated initiation times in Drosophila \cite{boettiger2009synchronous}.  

We also described the number of mRNA produced over a given period of time at one choice of $b$
(the probability the GTF scaffold dissociates before the return of the next polymerase).
Setting $b=0.8$, we found the distribution of the time delay between 
the beginning of the production of subsequent transcripts under each model.
Using this distribution, we simulated the number of mRNA produced during a 600 minute period 
in 2000 independent cells, under both the IR and the ER scheme 
(for the common vector of rate parameters listed above).
The resulting distributions of mRNA numbers are shown in figure \ref{fig:sim_results}B and C.
To depict the amount of variability this represents,
figures \ref{fig:sim_results}D and E show a cartoon of the results --
for each cell pictured, we sampled a random number of mRNA as above, which are shown red dots randomly scattered within the cell.
To emphasize the variability, we then colored cells blue that have less than two-thirds the mean mRNA number 
and colored cells red that have more than three halves the mean mRNA number.  

In this example, $\eta$ is 2.8 times larger in the ER model than in the IR model,
so these simulations also give a sense of how a given ratio of transcript count variabilities $\eta$ for the two schemes
corresponds to a difference in cell-to-cell variability of transcript counts, 
a topic we explore in more detail below.    

\subsection*{Effects of regulation scheme on expression timing}
Our predictions for the time of expression and the number of transcripts in the previous subsection
depended on the chosen parameter values 
such as the association rate of different GTFs and the average burst size of the gene expression.  
The values of such parameters can, for the most part, be only very approximately estimated.
Moreover, they may be expected to vary considerably between different genes and different species.

Since a single vector of parameters simultaneously specifies our models for the two regulation mechanisms,  
we can systematically explore all possible combinations of promoter strength and enhancer activation rates
and ask in each of these cases how the two mechanisms compare
in terms of speed, synchrony and variability in transcript counts.  

To compare the two kinds of regulation of the model in figure \ref{fig:picassembly}, 
we sampled 10,000 random vectors of transition rates and substituted them into our analytic expressions for $\mu_\tau$, $\sigma^2_\tau$, and $\eta$, 
with each rate chosen independently and uniformly between 0 and 1 (we could also
have used a regular grid of parameter vectors). 
Since we will use ratios of the relevant quantities to compare models,
and these ratios are all invariant under a common linear rescaling of time,
the fact that all rates are bounded by 1 is no restriction -- we are effectively sampling over {\em all} of 
parameter space.
(For instance, the ratio of mean expression times of the two models does not change after multiplying every rate parameter by 100.)
Furthermore, independent draws of new sets of 10,000 parameter vectors and substitutions give nearly identical results,
confirming that our results are not sensitive to the specifics of the sample.
Additionally, discarding parameter vectors for which the enhancer dynamics are significantly slower than for the promoter chain
(i.e.\ $k_{ab}$ or $k_{ba}$ is smallest)
does not qualitatively change any of the results, validating our treatment of the enhancer chain 
when analyzing the ER scheme. 

In figure \ref{fig:modelresults}A--C we plot the histogram of $\log_2$ ratios 
for the mean delay, variance in delay, and transcript count variability
for the 10,000 randomly selected parameter combinations sampled uniformly across parameter space.  We found that at all sampled choices of rate parameter, and therefore in the vast majority of parameter space,
the time to the first transcription event after induction is smaller
and less variable (i.e.\ more synchronous) for elongation regulation than for initiation regulation
in the realistic model of figure \ref{fig:picassembly}.  Thus, both the experimentally reported speed \cite{yao2007} and synchrony \cite{boettiger2009synchronous} for elongation regulated genes 
can be expected purely from effects of regulation topology without invoking changes in promoter strength or in the composition of the PIC.

We emphasize that this conclusion is still consistent with the possibility that a particular
initiation regulated gene is expressed in a more synchronous pattern or with more rapid kinetics than some other elongation regulated gene:
it is only necessary that the rate parameters are also sufficiently different.  
However, for the fixed set of rates associated with a given gene, 
the network topology of the ER scheme always improved synchrony and speed in our model of transcription relative to the corresponding IR scheme
for the parameter vectors we sampled.

There is a plausible intuitive explanation for why elongation regulation is almost always faster than initiation regulation (figure \ref{fig:modelresults}A).  
When the regulation acts downstream, there are multiple paths which the system can take to before it reaches the regulated step -- 
(i.e.\ either the enhancer can reach the permissive state first or the polymerase can load), 
as illustrated for the simple model in figures \ref{fig:markovmodel}A and B.  
The system moves closer to the endpoint with whichever happens first, whereas the IR regulated scheme must wait for enhancer activation before proceeding.  
The combination of this intuition and
our strong numerical evidence suggests a provable global inequality.
However, recall that for the toy model IR is faster over about 6\% of parameter space,
and one can reduce the realistic model to the toy model by making appropriate transitions very fast.
For example, for the toy model the choice of parameters
\begin{displaymath}
[k_{ab}, k_{ba}, k_{12}, k_{21}, k_{23}, k_{34}] =[ 1,  1,  .1,   .1,   .1, .0001]
\end{displaymath}
leads to a 5 fold increase in speed of the IR scheme relative to the ER scheme.  
This allows us to find parameter vectors where IR is faster than ER for the realistic model, for instance,
\[
\begin{split}
& [k_{ab}, k_{ba}, k_{12}, k_{21}, k_{23}, k_{32}, k_{24}, k_{42}, k_{35}, k_{53}, k_{45}, k_{54}, k_{56}, k_{65}, k_{67}, k_{78}] \\
& \quad = 
[.1, 1,   .01,   .01,   .01,   .01,   .01,   .01,   .01,   .01,   .01,   .01,   .01,   .01,   .01, .0001] \\
\end{split}
\]
produces in the realistic model a 10 fold increase in speed for the IR scheme relative to the ER scheme.  
However, such reversals of the typical ordering must occur over less than one ten-thousandth of parameter space.
The fact that the typical ordering is not universal and hence not the consequence of some 
analytically provable domination of one model by the other demonstrates 
the necessity of our numerical exploration of parameter space.

\subsection*{Effect of regulation scheme on mRNA concentration}

The effect of the regulatory scheme on the variation in the total amount of expression among cells 
is perhaps the most interesting and also experimentally untested consequence 
of regulating release from the paused state.  
As discussed above, we compute a factor $\eta \approx (\sigma_{N(t)}^2 / \mu_{N(t)}^2 )t$ for each scheme
and compare the schemes by examining the ratio of the resulting quantities.   
If the ratio $\eta_{IR}/\eta_{ER}$ is larger than one at a particular set of parameter values,
a population of cells using the IR scheme with those rate parameters
will show more variability in mRNA concentrations between cells
(relative to the average over all cells) than if they were using the ER scheme with the same rate parameters.
In this case, we say that the ER scheme is more {\em consistent} than the IR scheme.

\begin{figure}[!ht]
  \begin{center}
    \includegraphics[width=\textwidth]{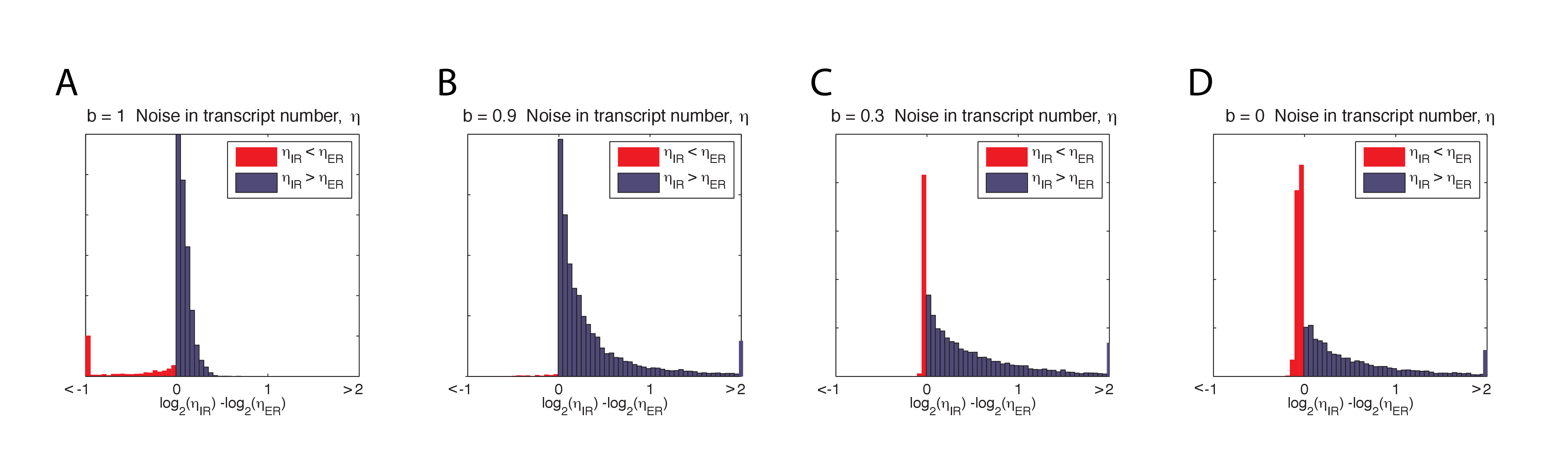}
  \end{center}
  \caption{{\bf Effect of scaffold stability} for variation in transcript number. 
  {\bf (A)}  $\log_2$ ratio of transcript variability, $\eta$, between the IR and ER model when all subsequent polymerases engage an assembled scaffold $b=1$.   
  Extreme values that would be off the edge of the graph are collected into the outermost bins.  
  {\bf (B)} As in (A) when $b=0.9$, note the ER scheme is more often substantially more coordinated, though a few parameters still make the IR scheme the more coordinated by a smaller margin.   
  {\bf (C)}   $b = 0.3$.   {\bf (D)}  $b=0$.
  }
  \label{fig:scaffold_effect}
\end{figure}

We explored the logarithm of this ratio 
(equivalently, the difference of the logarithms of the respective $\eta$ quantities) 
at four different values of $b$ (the probability the scaffold does not disassemble; see figure \ref{fig:picassembly});
several of the resulting distributions are shown in figure \ref{fig:scaffold_effect}.

When the complex is very stable, so that all polymerases find a preassembled scaffold to return to ($b=1$, figure \ref{fig:scaffold_effect}A), 
the ER scheme is more consistent for most rate parameters, but the differences are small.  
In fact, in nearly all cases at which $\eta$ differs by a factor of at least 2, the IR scheme is the more consistent.

When the scaffold is still stable but less so ($b=0.9$, figure \ref{fig:scaffold_effect}A; mean burst size 10),
the ER scheme still almost always produces more consistent numbers of transcripts among cells than the IR scheme,
and the differences are much larger.
If the scaffold is less stable ($b=0.3$, figure \ref{fig:scaffold_effect}C; mean burst size 1.4), 
the ER scheme is still more often more consistent than the IR scheme.

When we consider the simplest case with no bursting ($b=0$, figure \ref{fig:scaffold_effect}D), 
the ER scheme produces less variation in total transcript (smaller $\eta$) for most of parameter space.  
Moreover, the distribution is strongly skewed to the right, 
to the extent that for the 20\% of parameter space where there is more than a 
1.5 fold difference between the two regulatory mechanisms the ER scheme is always less variable.

We have found that, regardless of the value of $b$, the ER scheme is more consistent over most of parameter space.
However, for that difference in consistency to be substantial, $b$ must not be too close to 1.
This is at first surprising, because if the scaffold remains assembled, 
so that the chain returns to state 5 of figure \ref{fig:picassembly},
an IR scheme seems to have a clear ``advantage'' -- it does not have to wait for the enhancer to arrive,
whereas the ER scheme does, and one might expect that this added stochastic event would only increase variability.

Consideration of how each chain depends on its starting state suggests an intuitive explanation for this difference.
The IR scheme differs more in the amount of time it takes to reach the 
synthesis state when started with or without a scaffold (state 5 or state 1)
than does the ER scheme.  Intermediate values of $b$ allow the possibility of some cells making many bursts 
by reverting to state 5 after each synthesis while other cells make dramatically less by reverting to state 1 after each synthesis.   In contrast, under the ER regulation scheme, cells that start again from state 1 or from state 5 have relatively more similar synthesis times, and thus relatively less variation.  The similar synthesis times result from the fact that ER is faster starting from state 1, for the reasons discussed above, and slower than IR when starting from state 5, because of the extra regulatory step before synthesis.  Consequently, an ER scheme reduces the noise associated with very stable transcription scaffolds (see \cite{pedraza-paulsson, Thattai2001, tkavcik2008rin}
for a discussion of this noise).

\subsection*{Pertinent properties of elongation regulation}

To further understand why elongation regulation results in faster, more synchronous, and more consistent gene expression over a wide range of parameters we investigated alternative post-initiation regulatory schemes.  This allows us to explore  how changing certain properties of the model of PIC assembly (the promoter chain)
will affect the results: 
Is the difference large because there are many steps between the IR step and the ER step,
or is it because there is no allowed transition leading backward out 
of the state immediately before the regulated step?
To explore these questions, we made modifications to the toy model of figure \ref{fig:markovmodel}
which we are able to analyze without the assumption of enhancer equilibrium.

First note that, as is shown in figure \ref{fig:markovmodel}C, 
the ER model is still faster, less variable, and more reliable (smaller $\mu$, $\sigma$, and $\eta$) than the IR model
over approximately 95\% of parameter space.
(It is also reassuring that the results are so similar to those for the more realistic model.)

We performed the same analysis after adding a reverse transition from state 3 back to state 2 (see figure S3A-B).
The results are shown in figure S3C,
and demonstrate that there is strikingly little difference between the two models of regulation.
This suggests that the absence of a backwards transition from the state immediately preceding the regulated transition
is an important factor in producing the differences between the models we observed above.
In the ER scheme of figure \ref{fig:markovmodel}, PIC assembly becomes ``caught'' in state 3, awaiting arrival of the enhancer.  (Similarly, the ER scheme of figure \ref{fig:picassembly} gets ``caught'' in state 7).  
After adding a transition $3 \to 2$, PIC assembly may run up and down the chain many times before it is in state 3 
at the same time the enhancer is in the permissive configuration,
and this counteracts any benefits in speed or reliability that may have been gained otherwise.
(It is not obvious that this will happen: 
the ER scheme of figure S3B still has ``more routes'' from state 1A to state 4 than the IR model,
so it may run counter to intuition that the IR model could be so often faster.)
This furthermore suggests that regulating after a state in which PIC assembly is ``caught'' reduces variation --
some polymerases may run from state 1 to 8 smoothly and fire very quickly, 
while others may go up and down the assembly process 
many times before they actually escape the promoter and make a transcript 
(as is suggested by the data of Darzacq et al.\ \cite{darzacq2007}), and this
will substantially spread out the times at which the first transcript is created.  

We also investigated the case in which the $2 \to 3$ transition is regulated
and observed a similar pattern -- see figure S3D-F.  This investigation supports the
intuition that it is the stability of the paused state, not simply the parallel assembly of enhancer complex and promoter complex, that is most important in understanding the different behavior of the two regulatory schemes.  It also suggests that these differences should be specific to genes that are regulated through paused (as opposed to poised or stalled) polymerase.

\subsection*{Sensitivity analysis}

Small variations in rate parameters between cells will occur if the number of TF or Pol II molecules is small,
so it is of interest to investigate how robust the properties of each regulation scheme are to such variation
and which jump rates affect each scheme the most.  
To measure this sensitivity, we compute the gradient of a quantity of interest (e.g.\ the mean induction speed)
with respect to the vector of jump rates, square the entries, and normalize so that the entries sum to one,
giving a quantity we refer to as {\em relative sensitivity} that is analogous to the ``percent variation explained'' in classical analysis of variance.  
Our analytic solutions for the quantities of interest make this computation possible.
For example, let $m(\mathbf r)$ denote the mean transcription time of the chain when the vector of transition rates is $\mathbf r$.  
Then, the relative sensitivity of $m$ to each rate $r_i$ is
$(\partial_{r_i} m(\mathbf r))^2 / \sum_j (\partial_{r_j} m(\mathbf r))^2$.
The larger this quantity is, the larger is the relative effect a small change in $r_i$ has on $m$.

\begin{figure}[!ht]
  \begin{center}
    \includegraphics[width=4in]{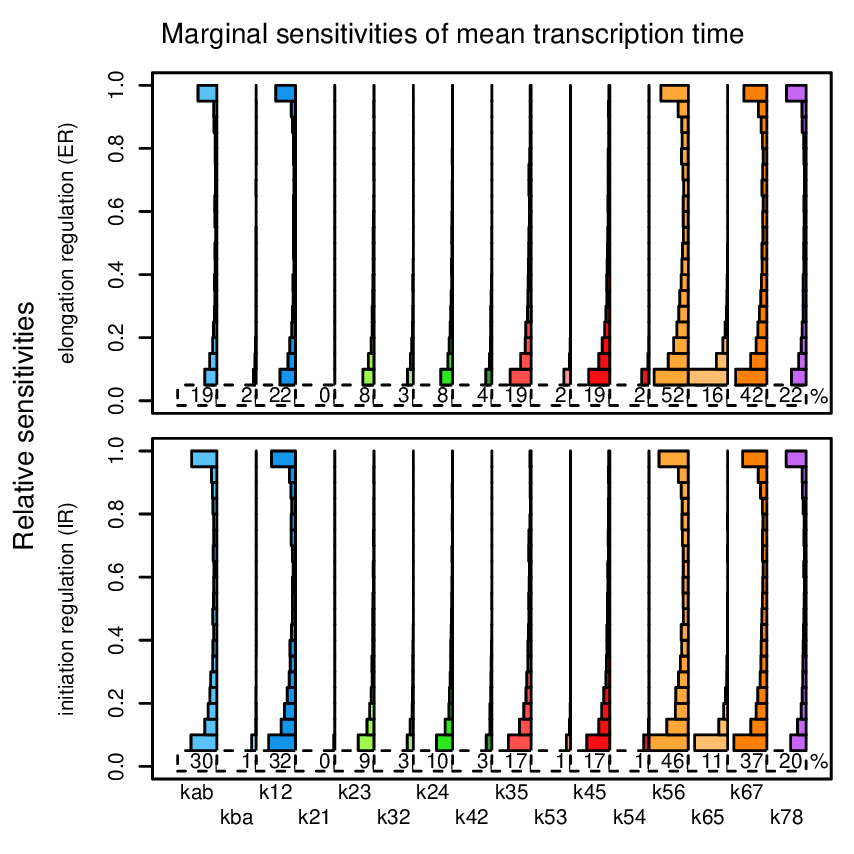}
  \end{center}
  \caption{{\bf Sensitivity Analysis} for mean expression time. 
    Histograms of the marginal distributions of relative sensitivities for both the ER and IR schemes,
    across uniform random samples from parameter space, as described in the text.
    The smallest bin of the histogram (values below $.05$) is disproportionately large, and so is omitted;
    shown instead is the percent of parameter space on which the relative sensitivity is at least $.05$.
    Note that often only a single parameter dominates (many sensitivities are near 1), 
    that many parameters are almost never influential,
    and that ER and IR are similar except for the addition of sensitivity to $k_{ab}$.
  }
  \label{fig:sensitivity}
\end{figure}

To explore the sensitivity across parameter space, 
we computed relative sensitivities for each of the three system properties to all 16 parameters
at each of the 10,000 random vectors of transition rates described above.
Each of the system properties showed surprisingly similar sensitivity profiles, 
so we only discuss the results for the mean time to transcription.
Marginal distributions of sensitivity of mean time to transcription to each parameter are shown in figure \ref{fig:sensitivity}.
Corresponding plots for the variance of transcription time and for transcript count variability are shown in figures S4 and S5.

As one might expect, for a given parameter vector the parameters to which the behavior 
of the models are most sensitive are generally those that happen to take the smallest value (and are thus rate-limiting):
for each parameter vector, we recorded the sizes of the two parameters with the highest and second highest sensitivity values 
and found that their sample means were $0.147$ and $0.296$, respectively (whereas the sample mean of
a typical parameter value will be very close to $0.5$).
However, just how small a given transition rate must be before it controls the system properties 
depends on where the corresponding edge lies in the topology of the network.  
As shown in figure \ref{fig:sensitivity}, 
some parameters are relatively important throughout a large region of parameter space in both the ER and IR schemes, 
while others only dominate the response of the system in a small portion 
and some never appear.

Two further observations are evident from this analysis.  
First, we see which transitions in the process of activating the gene are most sensitive to small fluctuations 
(due to small number of TF molecules or changes in binding strength).  
As is apparent from figure \ref{fig:sensitivity}, just 4 of the 16 promoter chain jump rates dominate the sensitivity,
and these are the same for both IR and ER schemes ($k_{12}$, $k_{56}$, $k_{67}$, and $k_{78}$). 
The relative importance among those 4 jump rates depends on the position in parameter space, 
primarily through their relative sizes.
Furthermore, although the ER and IR schemes have otherwise similar sensitivity profiles, 
the IR scheme is additionally sensitive to variation in the rate of enhancer--promoter interactions, $k_{ab}$.
As this interaction between potentially distant DNA loci is likely rate-limiting for gene expression, 
the robustness of the elongation regulated scheme to fluctuations of this rate 
may provide a further explanation for why elongation regulated genes appear to exhibit considerably more synchronous activation.  
It suggests additionally  that the rate of enhancer--promoter interactions is under more selective pressure for IR genes, 
where it has a large effect on their expression properties, than it is for ER genes, 
which may exhibit very similar expression properties despite having different enhancer interaction rates.

Second, we also observe that the complex assembly steps which may occur in arbitrary arrival order, 
namely the recruitment of TFIIE or TFIIF (governed by the jump rates $k_{23}$, $k_{24}$, $k_{35}$, and $k_{45}$) 
are considerably more tolerant to stochastic variation than sequential assembly steps such as the initial recruitment of the polymerase ($k_{12}$), 
the arrival of the last component of the complex, TFIIH ($k_{56}$), or promoter escape ($k_{67}$).  
Although between--cell variation in the total concentration of these intermediate, non-sequential binding factors will affect their binding rate parameters,
it will not greatly change properties of the time to expression, thus suggesting an additional benefit of ER.
This observation leads to the conclusion that the regulatory processes controlling the concentration of factors arriving in arbitrary order
and the binding affinities of such factors may be under less evolutionary pressure than 
the corresponding quantities for factors associated with other transitions.

\section*{Discussion}

Speed, synchrony, degree of cell--to--cell variability, and robustness to environmental fluctuations are important features of transcription.  
They are properties of the system rather than of a particular gene, DNA regulatory sequence, or gene product taken in isolation,
and optimizing them can, for instance, reduce the frequency of mis-patterning events that arise due to the inherently stochastic nature of gene expression.  
Understanding how these properties emerge, the mechanism by which they change, 
and the tradeoffs involved in optimizing them all require tractable models of transcription.

Through a study of stochastic models of transcriptional activation, we demonstrated that the increased speed and synchrony of paused genes, 
reported by Yao et al.\ \cite{yao2007} and Boettiger et al.\ \cite{boettiger2009synchronous} respectively, 
are expected consequences of the elongation regulation shown by such genes.  
We also predicted that ER genes produce more consistent numbers of total transcripts than IR genes. 
This hypothesis can be tested directly using recently developed methods (see \cite{bates2008,raj2009singlemolecule} for reviews and the
Supporting Information for more details).

We furthermore explored what aspects of ER make this possible.
From an examination of the effect of scaffold stability we proposed that elongation regulation should reduce the noise-amplifying nature of bursty expression.  
By investigating alternative models of post-initiation regulation, we also determined that our predictions 
depend critically on the stability of the transcriptionally engaged, paused polymerase, 
and would not be expected from polymerases cycling rapidly on and off the promoter 
(i.e.\ polymerase stalling).  

Our investigation required us to introduce a general probabilistic framework for analyzing system properties of protein--DNA interactions.  
Stochastic effects, resulting from molecular fluctuations, are increasingly understood to play important roles in gene control and expression 
(see \cite{raj2008review} for a review). 
We can now determine quantitatively how an element's location in a network affects the general properties of that network, 
even when the rate constants and concentrations of the network components are unknown.  
In particular, we quantified the extent to which system properties are sensitive
to each rate parameter, something which might predict the evolutionary constraint on that component.
Most previous approaches to the analysis of protein--DNA interactions have either relied on simulations, which require some knowledge of numerical rate values,
or use the fluctuation--dissipation theorem assuming the system is near equilibrium and the noise is small.  
Our methods avoid the limitations of those approaches
and also make analysis of realistic models, as done in \cite{coulon2010spontaneous}, significantly more feasible.

Finally, our approach is not restricted to investigating the assembly of transcriptional machinery,
but may also prove useful in studying stochastic properties of a variety of regulatory DNA sequences (such as enhancers). 
Different assembly topologies, such as sequential versus arbitrary association mechanisms for the component TFs \cite{hager2009transcription},
may account for some of the observed differences in sensitivities and kinetics between otherwise similar regulatory elements.  
As new technologies allow better experimental determinations of these mechanisms,
a theoretical framework within which one can explore their potential consequences will become increasingly important. 


\section*{Acknowledgments}

We thank Graham Coop, Mike Levine, George Oster, Dan Rokhsar, Ken Wachter, Michael Cianfrocco and Teppei Yamaguchi for helpful discussions and comments on the manuscript.



\end{document}